\documentclass[screen,nonacm]{acmart}

\AtBeginDocument{%
  \providecommand\BibTeX{{%
    \normalfont B\kern-0.5em{\scshape i\kern-0.25em b}\kern-0.8em\TeX}}}

\begin{document}

\title{Detox Browser - Towards Filtering Sensitive Content On the Web}

\author{Noble Saji Mathews}
\email{ch19b023@iittp.ac.in}
\orcid{0000-0003-2266-8848}

\author{Sridhar Chimalakonda}
\email{ch@iittp.ac.in}
\orcid{0000-0003-0818-8178}
\affiliation{%
  \institution{\\Research in Intelligent Software \& Human Analytics (RISHA) Lab,\\Department of Computer Science and Engineering,\\ Indian Institute of Technology Tirupati}
  \country{India}
}

\renewcommand{\shortauthors}{Noble and Sridhar}

\begin{abstract}
The annual consumption of web based resources is increasing at a very fast rate, mainly due to increase in affordability and accessibility of the internet. Many are relying on the web to get diverse perspectives, but at the same time it can expose them to content which is harmful for their mental well-being.
Catchy headlines and emotionally charged articles increase the number of readers which in turn increases ad revenue for websites. When a user consumes a large quantity of negative content 
, it adversely impacts the user's happiness and has a significant impact on his/her mood and state of mind. Many studies carried out during the COVID-19 pandemic has shown that people across the globe irrespective of their country of origin have experienced higher levels of anxiety and depression. 
Web filters can help in constructing a digital environment which is more suitable for people prone to depression, anxiety and stress.  Significant amount of work has been done in the field of web filtering, but there has been limited focus on helping Highly Sensitive Persons (HSP's) or those with stress disorders induced by trauma. Through this paper, we propose \textit{Detox Browser}, a simple tool that enables end users to tune out of or control their exposure to topics that can affect their mental well being. The extension makes use of sentiment analysis and keywords to filter out flagged content from google search results and warns users if any blacklisted topics are detected when navigating across websites.
Demonstration of \textit{Detox Browser} can be found here\footnote{\url{https://youtu.be/GZy_1jI9Uz4}}.
\end{abstract}



\maketitle

\section{Introduction}
\label{intro}

Due to ease of use and convenience, more and more people rely on web content from search engines and social platforms for reading, based on their topics of interest \cite{de2014seeking}. Rise in popularity of easily accessible online content has increased the possibility
of incidental exposure to news - a situation in which people consume content even when they actually aim to do something else \cite{essay81715}. 
In case of content which deals with situations of collective trauma like school shootings and acts of terrorism can cause anxiety \cite{luszczynska2009self}. Based on type content and quantity of content to which a person is exposed, a person can become more vulnerable to trauma related mental health problems over time \cite{holman2020media}. According to Sharma et al's study of mental health scenario in India during the COVID-19 pandemic, sensationalized news-stories increase anxiety so it is desirable to avoid them \cite{sharma2020indians}. 
f

In a study by Olagoke to evaluate the psychological impact of exposure to COVID news in the main stream media, it was found that young educated adults who perceive themselves to be vulnerable to COVID are more prone to depression \cite{olagoke2020exposure}. 
Unlike the individualistic culture of countries like the USA, India has a collectivistic culture where family bonds are considered more valuable in general and a greater share of people live in joint families \cite{migliore2011relation}. In such a situation, COVID related news may increase the fear that family members in high risk categories can contract COVID which in turn may lead to a higher level of depression, anxiety and stress \cite{wang2020immediate}. 
Recently, there has been a push for provision of mental health services through various information and communication technologies \cite{figueroa2020need}. 
But given the stigma associated with mental health, the general public would be reluctant to involve a third party in matters concerning their mental health \cite{shidhaye2013stigma}. Hence, in order to escape the infodemic \footnote{\url{https://www.who.int/health-topics/infodemic}} which came along with the COVID pandemic, people may need tools that can help control the content they view


Fear, depression and stress are some of the psychological symptoms reported by people who had to quarantine themselves due to COVID \cite{brooks2020psychological} and people with prior trauma tend to spend more time on news related to the pandemic which in turn may lead to increase in the severity of post-traumatic stress disorder (PTSD) \cite{solomon2021overwhelmed}. Thus it essential for people with prior trauma to filter out content which can act as PTSD triggers.  It is quite possible that there are people who see news on politics as a stress inducer and would like to avoid such news to maintain their mental stability. According to Adreas and Mathes \cite{nanz2020learning}, this exposure to news can be distracting and can cause people to waste time contemplating over frivolous news. Frank \& Pero \cite{marcinkowski2020incidental} express that such exposure to news can cause aversion to it. 

To solve these problems we conceptualised and designed the \textit{Detox Browser} which enables a user to filter out content which he/she might feel is adversely affecting their mental well-being. In addition to default filters, the extension enables a user to customize the browsing experience as per their personal requirements. The extension analyzes sentiment, categorizes content and removes blacklisted topics to prevent the user from being overwhelmed with information which can possibly traumatize him / her. The extension achieves this through simple keyword-based checks followed by a detailed analysis. Even though it is mainly aimed at google search results, it also checks websites for topics blacklisted by the user. The extension is in open beta and its features are being continuously improved. 

\section{Related Work}
\label{related work}

According to the study by Wu \& Li \cite{10.3389/fpsyg.2021.564284}, exposure to negative news related to COVID-19 can potentially cause depression. Given the harmful effects many have recommended spreading awareness about various mental health disorders and explored using tele-medical and e-Health interventions for treatment of such cases as described by Fonseca et al \cite{fonseca2021using}. Currently, the medical infrastructure is highly strained and available infrastructure to treat such diseases is limited, so it is important to focus on prevention rather than treatment. 
\cite{rauschenberg2021evidence}.

URL based content filtering is one of the most commonly available software solutions to filter web content, but it requires a database of URLs which is cumbersome to maintain \cite{akbas2008next}. 
Google has an inbuilt feature “safe search” which removes results which are age inappropriate. Though Google’s inbuilt safe search function filters out obscene content, but it cannot be customised according to a user’s requirement. Also given the nature of content of information which has to be filtered out, Google’s safe search is not enough. The functionality can further be improved with chrome extensions like Profanity Filter \footnote{\url{https://chrome.google.com/webstore/detail/advanced-profanity-filter/piajkpdbaniagacofgklljacgjhefjeh?hl=en}}, Safe Words\footnote{\url{https://chrome.google.com/webstore/detail/safe-words/mncnlkabidgflobdapkgmnlikkiilijj}} and many more which can remove profane words and even censor it with symbolic stand-ins for the purpose obfuscating these words \cite{suliman2017explicit}. These extensions focus on removing profane words but not on other types of content which the user might want to censor or might be sensitive to.

There have also been tools such as the Good News chrome extension \footnote{\url{https://chrome.google.com/webstore/detail/good-news/ekcmgkhakdlcondlgmadpiogjnlggpne}} which block news stories based on keywords and phrases, but this works only on the Google News website. Further to the best of our knowledge, there has not been much effort towards the use of sentiment analysis on the returned search results for methods to enable users to have a better control over their web browsing experience.

Machine learning has also played an important role in the field of web filtering. Tools have been developed to automatically remove malicious comments which can ruin the user experience of news readers \cite{sood2012automatic}. A similar approach was used to build NewsWeeder which uses user ratings to categorize and display news to the users liking \cite{lang1995newsweeder}. 

 
From the existing literature, we can see that significant work has been done in the development of abuse or profanity filters. However, not much has been done towards personalised search filtering based on the sentiment of the articles and in enabling users to have control over displayed content across the web. To address this, we propose \textit{Detox browser}. 


\section{Design and Development}
\label{dev}

\begin{figure}
    \centering
    \includegraphics[width = \linewidth, height=8.5cm]{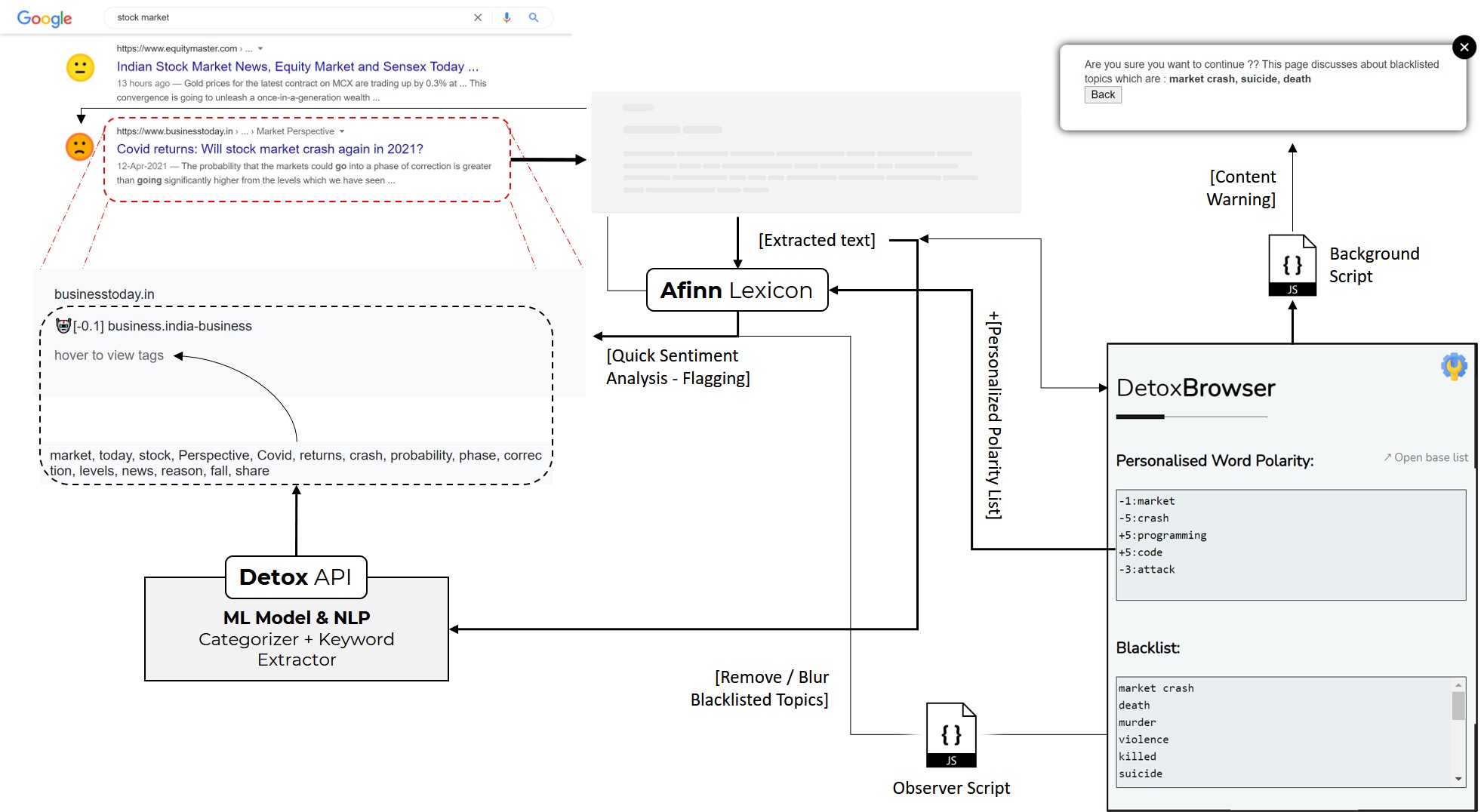}
    \caption{Workflow used by \textit{Detox Browser}}
    \label{fig:dev}
\end{figure}

\textit{Detox Browser} has been built as a chrome browser extension, primarily targeted at filtering google search results. It also supports profanity detection and content warnings for cross website navigation. The behaviour of the extension in each context can be modified via the settings and the sensitivity can be adjusted through the personalization options. The workflow utilised by the current version of the extension is shown in Figure \ref{fig:dev}.  

Once a google search is detected based on the URL, the extension extracts all HTML nodes from the page that have links in them. These extracted links are then used to obtain the closest parent nodes. They are categorised based on which of the predefined patterns they match. These patterns for categorization are obtained by manually analysing the selectors for the required components of the webpage being analysed. Currently, these patterns have been extracted for google search and the extension supports normal search results, featured stories, news and videos while ignoring special elements and Wikipedia / dictionary results so that direct searches for a topic and its meaning are not blocked.
 
\begin{figure}
    \centering
    \includegraphics[width = \linewidth,height=8.5cm]{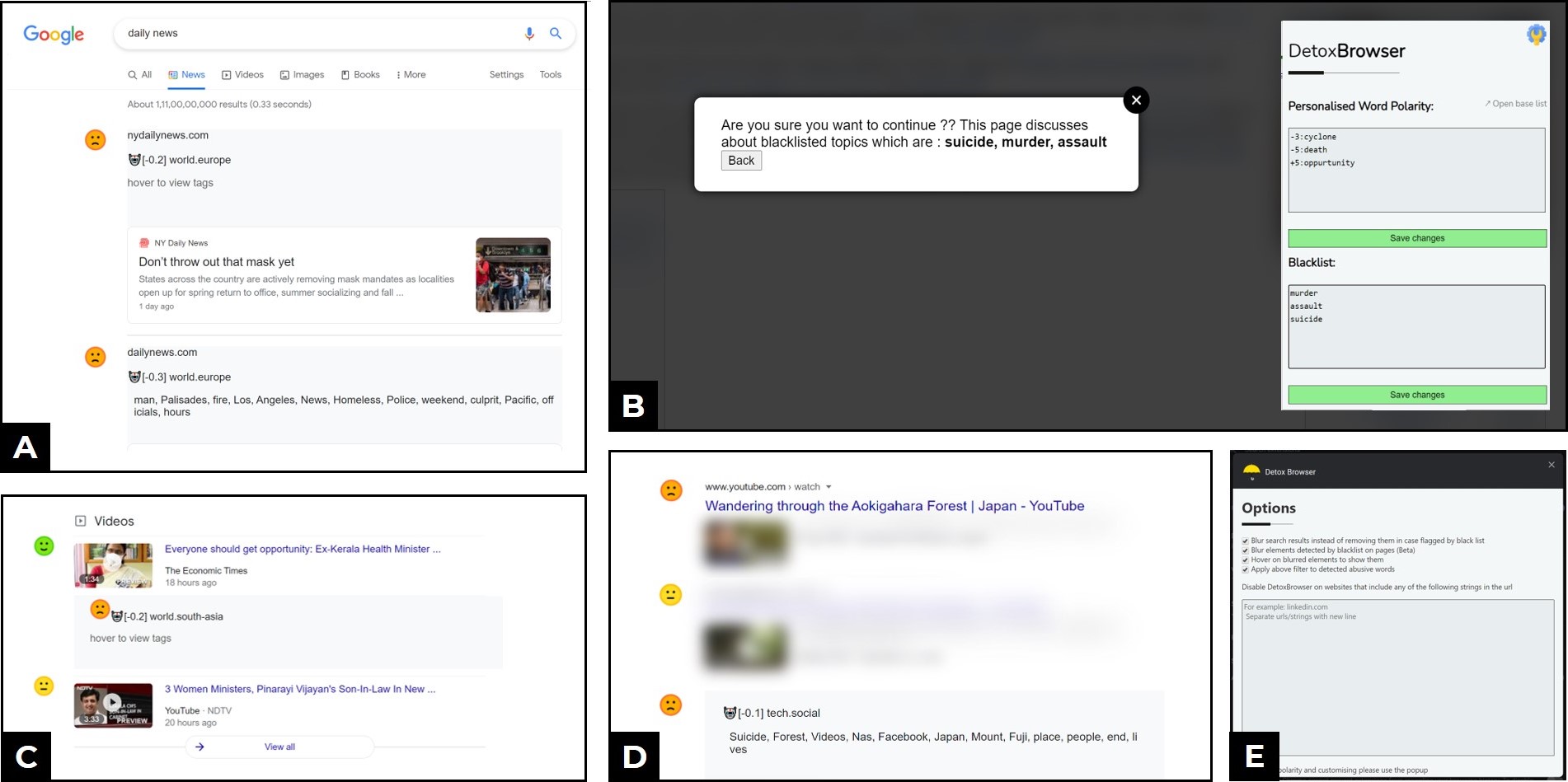}
  \caption{Screenshots of the extension in action, (A) Flagged results and tags on hover, (B) External Content Check and Personalising the extension, (C) Different sentiment scores, (D) Flagged content with blur enabled, (E) Extension Options }
  \label{fig:screens}
 \end{figure}

A mutation observer which is a built-in object that watches for any changes to a DOM element, ensures that any changes in the search page content are checked by the extension. From the parent nodes the text content is extracted by recursively traversing the Document Object Model (DOM). This text is then checked using a keyword based sentiment analysis based on AFINN
. Sentiment analysis can be done either through machine learning techniques or through a labeled set of data, to keep the size of the extension reasonable we went with the latter. Some of the popular lexicon based tools include VADER, SentiWordNet and AFINN. In our extension we have made use of AFINN as it has been shown to perform well in negative sample ratings \cite{al2020evaluating}.
The Afinn dataset contains words which are rated from -5 (negative) to +5 (positive), this allows for quick preliminary analysis in order to facilitate flagging of articles on page load based on if the value comes out to be negative. Flagged articles are replaced by a placeholder div that can be clicked to reinstate the swapped element. An emoji depicting strongly negative, negative, neutral, positive and strongly positive is appended to the left of the articles indicate the score obtained via lexical analysis as shown in figure \ref{fig:screens}[C].

For further checks, the extracted text is passed through a Multinomial Naive Bayes Classifier and Natural Language Processor. To keep the extension compact and for easy updates and improvements to these, they are deployed online and served through an API endpoint. The classifier is trained on a 20 Years Times of India Headlines data-set \cite{DVN/DPQMQH_2020} and categorises the articles into the top 50 categories from the data-set which has over 300 groups. 
The user can hover over content hidden by the extension to see the keywords generated through NLP and in case the user feels that it is a false positive then he/she can unhide it as shown in Figure \ref{fig:screens}[A] . The placeholder also mentions the domain name of the original article to help the user judge if he wants to see the article.
 
 The extension popup allows users to tweak the sensitivity to their liking. The polarity list allows specification of phrases along with a score from -5 to 5 which are used to override default values from AFINN. Further, the blacklist enables users to totally remove topics which they might not want to see. Depending on the options selected in the menu shown in Figure \ref{fig:screens}[E] the blacklist behaviour varies. The words / phrases in the blacklist are searched for using regex queries and if detected they are blurred by default with an option to remove on hover an example is shown in Figure \ref{fig:screens}[D]. For search results if the blur is disabled the extension will totally remove the result which contained the blacklisted keywords. This behaviour is also adopted to filter profanity and abusive language. For this, the profane-words package \footnote{\url{https://github.com/zacanger/profane-words}} is used which contains a large collection of profanity in English. The extension also packages a background script that warns the user if any of the blacklisted keywords are detected on a web site other than google search that the user visits with a popup as shown in Figure \ref{fig:screens}[B]. Through the options panel one may disable the extension on certain websites.

 
\section{Discussion and Limitations}

 
 Through \textit{Detox browser} we aim to  help improve mental health of the users by providing them the ability to have control on the content they see on the web. Through preliminary testing of the extension, we found that it tends to flag articles quickly. This trigger happy behaviour can be attributed to the lexical analysis and the limited amount of text in the title and description to judge the content of a search result. Over the course of development, we had tried analysing the entire content of the main article by loading it in the background, however, this approach was quite resource intensive, and hence was not pursued. The lexical analysis was mostly chosen because of its small bundle size and quickness. However, this causes a lot of false positives as it does not take into account the context based meaning of the statement. The categorizer is trained on Indian headlines hence is region specific to an extent. The sentiment analysis in place also works only in English and the tool does not support local languages.
 
 The tool is pending evaluation and a proper user study with volunteers with prior trauma, would be helpful in modifying the extension to cater to specific needs of such users. We also believe that our tool could be used by the general public, but we cannot comment on its efficacy yet. Another limitation of our approach is that we need to specify patterns for the elements to be analysed and replaced properly. This means that adding support for new websites requires manual effort and also that major updates to the way content is displayed on the target website can break the primary scripts. To take this into account we have kept checks in place that warn the user when the patterns no longer return results as expected.
 
  \section{Conclusion and Future Work}

 In this paper, we introduced \textit{Detox Browser}, a Google Chrome extension that filters search results as per the user's preferences. It also gives a popup warning if the content on any website is blacklisted by the user. 
 In the future we plan to extend the native scripts support to social media websites. Adding these direct scripts provide much more flexibility in ways we can control the content being delivered on the page. Taking into account dislikes in videos and image metadata for more features surrounding online media would help improve the tool as well. Further, since the Classifier and NLP model are deployed on the server side, we can keep updating the categorizer and NLP toolset to improve its accuracy. The categorizer could also be made to use more generic classes to account for use in multiple regions. Further expansion includes adding a secondary check with context-aware sentiment analysis to reduce false positives without the user's intervention in tuning the filter's sensitivity. In order to make it user friendly, we wish to introduce starter keywords in future versions so that casual users can slide through default sensitivity levels based on personalised word lists implemented and shared by the volunteers in the open beta.


\bibliographystyle{ACM-Reference-Format}
\bibliography{sample-base}










\end{document}